\documentclass[11pt]{article}
\usepackage{theorem}
\usepackage{latexsym}
\usepackage{amsmath}

\usepackage[dvips]{graphicx}
\usepackage{graphicx}

\newcommand{\newc}{\newcommand}

\newc{\be}{\begin{equation}}
\newc{\ee}{\end{equation}}
\newc{\bea}{\begin{eqnarray}}
\newc{\eea}{\end{eqnarray}}
\newc{\beas}{\begin{eqnarray*}}
\newc{\eeas}{\end{eqnarray*}}

\newc{\pardt}{\partial_{t}}
\newc{\pardxi}{\partial_{i}}
\newc{\pardts}{\partial_{t^{*}}}
\newc{\pardxis}{\partial_{i^{*}}}
\newc{\pardxj}{\partial_{j}}
\newc{\pardxk}{\partial_{k}}
\newc{\pard}{\partial}

\newc{\s }{\overline}
\newc{\sect}{\section}
\newc{\subs}{\subsection}

\newc{\defi}{\definition}
\newc{\prop}{\proposition}
\newc{\rem}{\remark}
\newc{\lem}{\lemma}
\newc{\exa}{\example}
\newc{\theo}{\theorem}
\newc{\coro}{\corollary}
\newc{\post}{\postulate}
\newc{\state}{\statement}

\begin{document}
\baselineskip0.5cm
\renewcommand {\theequation}{\thesection.\arabic{equation}}
\title{Vortex density waves and high-frequency second sound
in superfluid turbulence hydrodynamics}

\author{D.~Jou$^1$,
M.S.~Mongiov\`{\i}$^2$\thanks{Corresponding author.}  and M.
Sciacca$^2$}

\date{}
\maketitle
\begin{center} {\footnotesize $^1$ Departament de F\'{\i}sica, Universitat Aut\`{o}noma de
Barcelona, 08193 Bellaterra, Catalonia, Spain\\
$^2$Dipartimento di Metodi e Modelli Matematici Universit\`a di
Palermo, c/o Facolt\`{a} di Ingegneria,\\ Viale delle Scienze, 90128
Palermo, Italy}

PACS number(s): 67.40.Vs, 67.40.Bz, 47.27.2i, 05.70.Ln
\end{center} \footnotetext{E-mail addresses: david.jou@uab.es (D. Jou), mongiovi@unipa.it (M. S. Mongiov\`{\i}), micheles@math.unipa.it
(M. Sciacca)}

\begin{abstract}
In this paper we show that a recent hydrodynamical model of
superfluid turbulence describes vortex density waves and their
effects on the speed of high-frequency second sound. In this
frequency regime, the vortex dynamics is not purely diffusive, as
for low frequencies, but exhibits ondulatory features, whose
influence on the second sound is here explored.
\end{abstract}

Hydrodynamics of superfluid turbulence, characterized by a tangle
of quantized vortex lines \cite{D2,BDV-Springer}, aims to describe
the couplings between pressure and heat perturbations, and the
vortex density dynamics \cite{NL-SP57}-\cite{MJ1}. Such
hydrodynamics is a lively topic, with recent emphasis, for
instance, on nonlinear features such as the influence of intense
heat pulses on the vortex tangle \cite{KNN-LTP25}-\cite{Z-PRB74},
or in multi-scale formulations allowing to eliminate the fast
processes to derive evolution equations for the slow processes
\cite{NP-Cry45}. In the present paper we will focus our attention
on a different topic, namely, the behaviour of linear heat waves
and vortex density waves in the high-frequency domain. This has
not received much previous attention, but we think it is
worthwhile because, whereas at low frequencies the behaviour of
the vortex distortions is mainly diffusive, at high enough
frequencies, the vortex lines behave as an elastic medium, and are
able to propagate waves by themselves, i.e. they behave like a
viscoelastic medium: diffusive at low frequencies, elastic at high
frequencies. Their effect on second sound propagation is of much
interest to provide a suitable physical interpretation of the
experimental data based on second sound.

Collective vortex waves in rotating superfluids have been studied
in depth for a long time \cite{S-RMP59}, and the so-called
Tkachenko transverse elastic waves in the vortex arrays with
crystalline order arising in rotating cylinders have been
theoretically discussed since 1966. However, vortex density waves
in counterflow situations have not been studied, up to our
knowledge.

Usually, one considers homogeneous vortex tangles in counterflow
situation (i.e. under vanishing barycentric speed) and the
evolution of $L$, the {\it vortex-line density}, is assumed to be
the well-known Vinen's equation
\cite{D2,BDV-Springer,NF-RMF67,V-PRSLA240}

\[{dL\over  dt} =\sigma_L= A q L^{3/2}   - B  L^2,\eqno(1)\]
 where $q$ is the absolute value of the heat flux ${\bf q}$, $A$ and $B$
parameters linked to the dimensionless coefficients appearing in
Vinen's equation by the relations $A=\alpha_v/ \rho_s Ts$ and
$B=\kappa \beta_v$ [7], $\rho_s$ being the density mass of the
superfluid component, $T$ the absolute temperature, $s$ the
entropy, $\kappa=h/m$ the quantum of vorticity ($m$ the mass of
the $^4$He atom and $h$ Planck's constant, $\kappa \simeq 9.97$
10$^{-4}$cm$^2$/s).

This equation is used to describe homogeneous turbulence. When the
vortex line density is inhomogeneous, such an equation must be
generalized incorporating nonlocal terms. In \cite{MJ1}, a set of
evolution equation for $\epsilon$, ${\bf q}$ and $L$, with
$\epsilon$ being the energy density, was proposed with special
attention to their consistency with the second law of
thermodynamics through the Liu's procedure. Such equations were
\cite{MJ1}

\[
 \left\{
  \begin{array}{l}
    \rho {\dot \epsilon}  +
 \nabla \cdot {\bf q} =0, \\
    \dot {\bf q} + \zeta_0 \nabla T + \chi_0   \nabla L = -K L {\bf q}, \\
    {\dot L} + \nabla \cdot (\nu_0 {\bf q})=\sigma_L.
  \end{array}
\right. \eqno(2)\]

The coefficient $K$ is related to the Hall-Vinen friction
coefficient $B_{HV}$ as $K=\frac{1}{3} \kappa B_{HV}$, $\nu_0$ and
$\chi_0$ are suitable coefficients whose physical meaning we want
to explore and $\zeta_0$ is the ratio of the heat conductivity of
the fluid over the relaxation time of the heat flux, which is
given by $(KL)^{-1}$ . A further term proportional to $L^{3/2}
{\bf \hat{q}}$ may be added to the right-hand side of the equation
(2b) for the heat flux, describing a dry friction term. Here, we
will omit it for the sake of simplicity \cite{NF-RMF67,MJ1}. For
vanishing $\chi_0$ and $\nu_0$, the two first equations describe
second sound (temperature waves) with speed $V_2^2 =
\frac{\zeta_0}{\rho c_V}$, with $c_V$ the specific heat per unit
mass at constant volume, and the third equation reduces to Vinen's
equation (1). In the steady state, the second equation reduces to
\[
{\bf q} = - \frac{\zeta_0}{K L}  \nabla T -  \frac{\chi_0}{KL}
\nabla L. \eqno(3)\] The coefficient of the first term is just the
heat conductivity already mentioned, and the second term describes
a coupling between the vortex line density and the heat flux,
analogous in some way to the coupling between concentration
gradient and heat flux in usual fluids (the so-called Soret
effect). The term in $\nabla L$ could be related to the internal
energy density of the tangle, which is given by $\epsilon_V L$,
with $\epsilon_V = (\rho_s \kappa^2/4 \pi)\ln(c/a_0 L^{1/2})$ the
energy per unit length of the vortices.
 In more specific terms, it may be related to the tangle
contribution to the total pressure, as given in (10). Since in
this paper we are concerned with high frequency and short
wavelength perturbations, the contribution of $\nabla L$ will be
relevant, in contrast with usual situations at long wavelengths.

From (2) it is seen that when $\dot {\bf q}$ is neglected and the
system is isothermal, and in the linear approximation, one obtains
from (2b) and (2c) a reaction-diffusion equation for $L$ of the
form

\[ {\dot L} = \frac{\nu_0 \chi_0}{K L_0} \nabla^2 L + \sigma_L, \eqno(4)\]
whose diffusion coefficient $D$ is given by $D=\frac{\nu_0
\chi_0}{K L_0}.$

If, instead, $\dot {\bf q}$ is not neglected in (2b) and still in
the isothermal assumption -- for the sake of simplicity -- we obtain

\[ {\ddot L}+K L_0 {\dot L}= \nu_0 \chi_0 \nabla^2 L + K L_0
\sigma_L+ \dot \sigma_L. \eqno(5)\] At high frequencies, i.e. for
$\omega \gg K L_0$ and much higher than the inverse of the
characteristic time of the vortex destruction and formation as
described by $\sigma_L$, namely $\omega \gg 2BL_0-\frac{3}{2} A q
L_0^{1/2}$, (5) becomes

\[ {\ddot L} \approx \nu_0 \chi_0 \nabla^2 L, \eqno(6)\] that is, we get
vortex waves with speed \[ v_\infty^2=\nu_0 \chi_0. \eqno(7)\]
This outlines the physical features of the coefficients $\nu_0
\chi_0$ as their product is seen to be related both to the
diffusion coefficient and to the speed of vortex waves.

Elastic compressional waves in the vortex density obtained in (6)
may also be interpreted in a more mechanistic way. First of all,
we may interpret the term $\nu_0 \bf q$ in equation (2c) as $\nu_0
{\bf q} = L\textrm{v}_L$, with $\textrm{v}_L$ an effective
velocity of the tangle because the divergence term in the balance
equation (2c) may be interpreted as a flux of vortex lines moving
with peculiar velocity $\textrm{v}_L$. This effective velocity
will have an evolution equation of the type
\[
\rho_{eff}  \frac{\partial \textrm{v}_L}{\partial t} = -\nabla
p_V, \eqno(8)\] where $\rho_{eff}$ is an effective density of the
vortices and $p_V$ is the vortex contribution to the pressure. In
fact, vortices themselves do not have mass, but they have
indirectly associated an inertia  due to the mass of the
surrounding rotating fluid. By following Sonin (page 94 of Ref.
12) we estimate this mass per unit volume as $\rho_s L r^2$, with
$\rho_s$ the density of the superfluid component, which is the one
which participates in the rotation around the vortices of the
fluid and $r$ a characteristic radius of the zone in which the
superfluid is affected by the motion of the vortex; this will be
of the order of the average vortex separation, namely $L^{-1/2}$;
in this way, we have $\rho_{eff} \propto \rho_s$. This is the
reason that $\rho_s$ is the density which appears in the
coefficient $\epsilon_V$ appearing in the expression for the
vortex density of the tangle, as given in the paragraph below
equation (3).

Then, we may combine equation (8) with equation (2c), which at
high enough frequency, when the influence of the
production-destruction term $\sigma_L$ is negligible, yields
\[
\frac{\partial^2 L}{\partial t^2} = -L_0 \nabla \cdot
\frac{\partial \textrm{v}_L}{\partial t} =  \frac{L_0}{\rho_{eff}}
\nabla^2 p_V. \eqno(9)
\]
The total pressure of the turbulent superfluid has the form
(equation (4.16) of Ref. 7)
\[
p = p^* + \epsilon_V L,    \eqno(10)\] $p^*$ being the pressure of
the bulk superfluid and $\epsilon_V L$ the contribution of the
tangle, with $\epsilon_V$ the energy per unit length of the
vortices. Combination of (9) and (10) and taking into account that
$\rho_{eff} \propto \rho_s$, yields
\[
\frac{\partial^2 L}{\partial t^2} = \left(L_0
\frac{\epsilon_V}{\rho_{eff}}\right) \nabla^2 L. \eqno(11)\]

Then the velocity of the vortex waves will be
\[
v_L^2 = \left(L_0 \frac{\epsilon_V}{\rho_{eff}}\right) \propto
\frac{L_0 \kappa^2}{4\pi} \ln\left(\frac{c}{a_0 L^{1/2}}\right).
\eqno(12)\]

The combination $L_0 \kappa^2$ is similar to the combination
$\Omega \kappa$ appearing in the velocity of inertial waves in the
vortex array in rotating superfluids \cite{S-RMP59}, if one
replaces $L_0 = 2\Omega/\kappa$, $\Omega$ being the angular speed
of the rotating cylinder.

We now go to the diffusion coefficient $D$ corresponding to
equation (4). Since we have obtained an estimation of $\nu_0
\chi_0$ appearing in (6), we will take advantage of it to obtain
an expression for $D$. We get
\[
D = \frac{\nu_0 \chi_0}{K L_0} \propto  \frac{\kappa}{B_{HV} 4\pi}
\ln \left(\frac{c}{a_0 L^{1/2}}\right) \eqno(13)
\]

Indeed, it was known that $D \propto \kappa$ on dimensional
grounds and on some numerical simulation \cite{TAV}. Here, we have
a more explicit expression, based on a more microscopic model.
Unfortunately, our derivation cannot set precisely the
proportionality constant in (13) due to the ambiguity in the mass
density $\rho_{eff}$ associated to the vortices (more precisely,
to the superfluid associated to the vortices). The analysis of the
high-frequency regime may be indeed rewarding in new results which
are not available if the study is limited to the low-frequency
regime.

Up to here we have considered an isothermal situation. According
to (2), the behavior of ${\bf q}$ and $L$ under non-isothermal
conditions is connected to the behavior of the field $T$ in such a
way that the complete study of the system (2) is required. In
\cite{MJ1} two of us have already tempted to solve this more
complicated situation, but now to the light of which we have said
before, some more explicit conclusions may be done with a clearer
physical meaning in the high frequency regime, which complements
the information obtained at low frequencies.

As in \cite{MJ1}, expressing the energy in terms of the vortex line
$L$ and the temperature $T$ and linearizing the two contributions
$\sigma_L$ and $L {\bf q}$ around the stationary solutions, already
found in \cite{MJ1},

\[ {\bf q}={\bf q}_0=(q_{10},0,0), \quad
L=L_0=\frac{A^2}{B^2}[q_{10}]^2, \quad T=T_0({\bf x})=T^*-\frac{K
L q_{10}}{\zeta_0}x_1, \eqno(14)\] the system (2) assumes the
following form

\[
 \left\{
  \begin{array}{l}
    \rho c_V {\dot T} + \rho \epsilon_L {\dot L}+
 \nabla \cdot {\bf q} =0, \\
    \dot {\bf q} + \zeta_0 \nabla T + \chi_0   \nabla L = -K [L_0 {\bf q}+{\bf q}_0 (L-L_0)], \\
    {\dot L} +\nu_0  \nabla \cdot {\bf q}=-\left[2BL_0-\frac{3}{2}Aq_{10}L_0^{1/2}\right](L-L_0)+A q_{10}L_0^{3/2}(q_1-q_{10}).
  \end{array}
\right. \eqno(15)\]

Now, if we suppose the propagation of harmonic plane waves of the
form

\[ \left\{
\begin{array}{ll}
T=T_0({\bf x})+\tilde{T} e^{i(\bar{K}{\bf n\cdot x}-\omega t)}\\
{\bf q}={\bf q}_0+\tilde{{\bf q}} e^{i(\bar{K}{\bf n\cdot x}-\omega t)}\\
L=L_0+\tilde{L} e^{i(\bar{K}{\bf n\cdot x}-\omega t)},
\end{array}\right.
\eqno(16)\] where $\bar{K}=k_r+i k_s$ is the complex wave number,
$\omega$ is the frequency and ${\bf n}$ the unit vector in the
direction of the wave propagation, then we have the propagation of
waves (vortex waves and heat waves) which move at the same speed
$w_2=\frac{\omega}{k_r}$ given by the combination of the second
sound and vortex waves velocities. As in \cite{MJ1}, we are
assuming that the quantities oversigned by a tilde are small ones,
and whose product can be neglected.

Inserting (16) in the linearized system (15) we obtain the
following algebraic system for the quantities ${\tilde T}$,
${\tilde L}$ and $\tilde {\bf q}$

\[
 \left\{
  \begin{array}{l}
    -[\rho c_V]_0 \omega {\tilde T} -[\rho \epsilon_L]_0 \omega  {\tilde L}+
 {\bar K} {\bf {\tilde q} \cdot n} =0, \\
    (-\omega-i N_1) \tilde {\bf q} + {\bar K} [\zeta_0]_0 \tilde T {\bf n} + \left({\bar K}  [\chi_0]_0 {\bf n} -i
    N_3 \hat{{\bf q}}_0\right) \tilde L =0, \\
    (-\omega-i N_2) {\tilde L} + {\bar K} [\nu_0]_0  {\bf \tilde q \cdot n} +i N_4\tilde
    q_1=0,
  \end{array}
\right. \eqno(17)\] where
\[
N_1=KL_0, \quad N_2=2BL_0-\frac{3}{2}AL_0^{1/2}q_{10}, \quad N_3=K
q_{10}, \quad N_4=A q_{10} L_0^{3/2},
\]
 and the subscript $0$, which will be deleted from
now on, denotes quantities referring to the unperturbed states.

Now, let consider that the direction of the wave propagation is
collinear to the initial heat flux, i.e. ${\bf n}=(1,0,0)$, then
the system (17) becomes

\[
 \left\{
  \begin{array}{l}
    -\rho c_V \omega {\tilde T} -\rho \epsilon_L \omega  {\tilde L}+
 {\bar K} {\tilde q}_1 =0, \\
    (-\omega-i N_1) \tilde q_1 + {\bar K} \zeta_0 \tilde T + \left({\bar K}  \chi_0 -i
    N_3\right) \tilde L =0, \\
(-\omega-i N_1) \tilde q_2 =0, \\
    (-\omega-i N_1) \tilde q_3 =0, \\
    (-\omega-i N_2) {\tilde L} + \left({\bar K} \nu_0+i N_4\right)\tilde
    q_1=0.
  \end{array}
\right. \eqno(18)\] In the hypothesis of high-frequency waves,
which means $\omega \gg N_1$, $\omega \gg  N_2$ and $|\bar{K}|\gg
max\left(\frac{Kq_{10}}{\chi_0}, \frac{Aq_{10}
L_0^{3/2}}{\nu_0}\right)$, the previous algebraic system (18)
becomes

\[
 \left\{
  \begin{array}{l}
    -\rho c_V \omega {\tilde T} -\rho \epsilon_L \omega  {\tilde L}+
 k_r {\tilde q}_1 =0, \\
    -\omega \tilde  q_1 + k_r \zeta_0 \tilde T + k_r  \chi_0 \tilde L =0, \\
    -\omega {\tilde L} + k_r \nu_0  \tilde q_1 =0,\\
-\omega  \tilde q_2 =0,\\
-\omega  \tilde q_3 =0.
  \end{array}
\right. \eqno(19)\]

The above system has nontrivial solutions if and only its
determinant is zero, which corresponds to the following dispersion
relation

\[ w_2^2=V_2^2(1-\nu_0 \rho \epsilon_L)+v_\infty^2, \eqno(20)\] where
$w_2=\omega/k_r$ is the speed of the wave, $V_2^2=\zeta_0/\rho
c_V$ is the second sound speed in the absence of vortices and
$v_\infty^2=\chi_0 \nu_0$ is the speed of the vortex wave, which
we have found in (7). We recall that all three fields $T$, $L$ and
${\bf q}$ vibrate with the same speed $w_2$ given by (20) in such
a way that each field contributes to the vibrations of the other
two. If we try to read the relation (20) in terms of the second
sound, we note that the vortex vibrations modify this second sound
speed through the two contributions $-V_2^2\nu_0\rho \epsilon_L$
and $v_\infty^2$, the latter due to the presence of the vortex
waves and the former due to the reciprocal existence of two waves.
The same conclusion may be achieved reading the relation (20) in
terms of the vortex waves. The correction for the speed of the
second sound is not important, because $V_2$ is of the order of
$20$ m/s near $1,7$ K \cite{ZMW-IJHMT49,NP-Cry45}, whereas, for
$L_0 = 10^6 \textrm{cm}^{-2}$, and according to the estimation
(12), the speed of vortex density waves would be of the order
$0,25$ cm/s, much lower than $V_2$.

In the earlier analysis of the system (15), we have only
considered the terms of the equations in which $\omega$ and $k_r$
appear such as  we have also neglected the term $k_s$ relative to
the dissipation of the wave. Now, we assume that the quantities
$N_1$, $N_2$, $N_3$ and $N_4$ are coefficients small enough to
assume them as perturbations of the physical system. This is
reasonable at high-frequencies, since we have assumed that $\omega
\gg  N_1$, $\omega \gg  N_2$ and $|\bar{K}|\gg
max\left(\frac{Kq_{10}}{\chi_0}, \frac{Aq_{10}
L_0^{3/2}}{\nu_0}\right)$. Thus, the contributions of these
coefficients can modify the speed of the wave in a small quantity
$\delta$ and the imaginary part of the wave number, $k_s$.

Therefore, let assume that the speed of the wave has the following
expression

\[ w=\frac{\omega}{k_r}=w_2+\delta, \eqno(21)\] for which substituting it in
the dispersion relation, i.e. the equation obtained imposing that
the determinant relative to the system (15) vanishes, we obtain at
the lower order the relation (20) for the speed $w_2$. From the
next order follows that $\delta=0$, that is the perturbations due
to the coefficients $N_1$, $N_2$, $N_3$ and $N_4$ do not modify
the speed of the wave while they modify the coefficients $k_s$
related to the attenuation in the form

\[ k_s^{\|}=\frac{N_2
   \left(w_2^2-V_2^2\right)+w_2
   \left(\rho \epsilon_L N_4 V_2^2+N_1
   w_2+N_3 \nu_0 -N_4 \chi_0
   \right)}{2 w_2^3}. \eqno(22)\]
Anyway, this modification will be small, because $w_2^2-V_2^2$ is
small and the coefficients $N_i$ are also small in the situation
considered.

Now, we consider the case in which ${\bf n}$ and the initial heat
flux ${\bf q}_0$ are orthogonal, and in particular we choose ${\bf
n}=(0,0,1)$ \cite{MJ1}. Through this choice the system (15)
becomes

\[
 \left\{
  \begin{array}{l}
    -\rho c_V \omega {\tilde T} -\rho \epsilon_L \omega  {\tilde L}+
 {\bar K} {\tilde q}_3 =0, \\
        (-\omega-i N_1) \tilde  q_3 + {\bar K} \zeta_0 \tilde T + {\bar K}  \chi_0\tilde L =0, \\
    (-\omega-i N_2) {\tilde L} + {\bar K}\nu_0  \tilde q_3 +i N_4\tilde
    q_1=0,\\
    (-\omega-i N_1) \tilde q_1 -iN_3\tilde L =0, \\
    (-\omega-i N_1) \tilde q_2 =0.
  \end{array}
\right. \eqno(23)\] If now we make the same assumption as the
previous case, i.e. high frequencies $\omega$ and high wave number
$k_r$, the last system becomes

\[
 \left\{
  \begin{array}{l}
    -\rho c_V \omega {\tilde T} -\rho \epsilon_L \omega  {\tilde L}+
 k_r {\tilde q}_3 =0, \\
     -\omega \tilde q_3 +  k_r  \zeta_0  \tilde T  +  k_r   \chi_0  \tilde L =0, \\
     -\omega {\tilde L} +  k_r \nu_0   \tilde q_3 =0,\\
     -\omega \tilde q_1  =0, \\
-\omega\tilde q_2  =0.
  \end{array}
\right. \eqno(24)\] The dispersion relation of this system is
found by setting to zero the discriminant of the matrix associated
to (24), from which the same speed $w_2=\omega/k_r$ of the wave of
the previous case is obtained

\[ w_2^2=V_2^2(1-\nu_0\rho \epsilon_L)+v_\infty^2. \eqno(25)\]

Now, following the same procedure of the previous case, we
consider the quantities $N_1$, $N_2$, $N_3$ and $N_4$ as small
perturbations of the physical system in such a way the speed of
the wave is modified by a quantity $\delta$ and the imaginary part
cannot be negligible. Therefore, assuming that the new speed of
the wave has the form $w=\omega/k_r=w_2+\delta$, it is found the
relation (25) for the speed $w_2$, at the lowest order, and
$\delta=0$ and \[ k_s^\bot=\frac{N_2
   \left(w_2^2-V_2^2\right) +N_1 w_2^2}{2
   w_2^3},
\eqno(26)\] at the successive order. Note that also in this case
the perturbations do not modify the speed of the waves but only
the dissipative term $k_s$. From a comparison between the two
relations of $k_s$, (22) and (26), one may note that the
quantities $N_3$ and $N_4$ take part only in (22), that is, only
when the direction of the wave propagation is collinear to the
initial heat flux. In particular, one may write
$k_s^{\|}=k_s^\bot+\frac{\left(\rho \epsilon_L
V_2^2-\chi_0\right)N_4+N_3 \chi_0}{2w_2^2}$.

In summary, we have shown that at high enough frequencies the
dynamics of inhomogeneous vortex tangles shows a crossover from a
diffusive to propagation behavior. We have studied the
consequences of these contributions on the speed of high-frequency
second sound in (20) and (25) and on the attenuation of
longitudinal and transverse second sound in (22) and (26). The
origin of the longitudinal density waves predicted by the equation
(2) is to be found in the vortex contribution to the total
thermodynamics pressure of the system, as described in (10). This
is different from the transverse elastic waves known as Tkachenko
waves found in rotating vortex arrays. An evaluation of the speeds
shows that the speed of the second sound, according to (20) and
(25), will not be much influenced by the presence of the vortex
waves; in contrast, the speed of vortex waves, which would be
rather small in the absence of second sound, is much increased in
the presence of high-frequency second sound, because in these
circumstances they propagate at a common speed (20) or (25). The
results for the attenuation coefficient according to (22) and (26)
are also interesting, because it turns out that the attenuation of
second sound at high frequency will be very low. This is in
contrast with what happens at low frequency, or when the vortex
tangle is assumed as perfectly rigid, not affected by the second
sound, in which case the relative motion of the normal fluid with
respect to the vortex lines yields an attenuation which allows to
determine the vortex line density $L$ of the tangle
\cite{D2,BDV-Springer}. However, the wave character of vortex
density perturbations at high frequency makes that vortex lines
and the second sound become two simultaneous waves with a low
joint dissipation, in the first-order approach. Thus, from the
practical point of view, it seems that, at high frequency, second
sound will not provide much information on the vortex tangle
because the influence of the average vortex line density $L_0$ is
small both in the speed as in the attenuation. We have also
obtained an explicit expression for the vortex diffusion
coefficient (13) in terms of $\kappa$ and $B_{HV}$, which is
somewhat analogous to the Einstein expression for the diffusion
coefficient, but with the quantum $\kappa$ of the turbulence
instead of the thermal energy and the friction coefficient
$B_{HV}$ in the denominator. This may be useful for studies on
hydrodynamics of vortex tangles \cite{NL-SP57,G-PhyA183,MJ1}.

\subsection*{Acknowledgments} We acknowledge the support of the
Acci\'{o}n Integrada Espa\~{n}a-Italia (Grant S2800082F
HI2004-0316 of the Spanish Ministry of Science and Technology and
grant IT2253 of the Italian MIUR). DJ acknowledges the financial
support from the Direcci\'{o}n General de Investigaci\'{o}n of the
Spanish Ministry of Education under grant Fis2006-12296-c02-01 and
of the Direcci\'{o} General de Recerca of the Generalitat of
Catalonia, under grant 2005 SGR-00087. MSM and MS acknowledge the
financial support from MIUR under  grant "PRIN 2005 17439-003" and
by "Fondi 60\%" of the University of Palermo. MS acknowledges the
"Assegno di ricerca" of the University of Palermo.

\end{document}